\pdfoutput=1
\documentclass[11pt,a4paper]{article}
\usepackage{geometry}
\usepackage{graphicx}
\usepackage{color}
\usepackage[colorlinks=true,citecolor=blue,linkcolor=magenta,urlcolor=magenta]{hyperref}
\usepackage{subcaption}
\usepackage{amssymb}
\usepackage{amscd}
\usepackage{amsmath}

\begin{document}
\begin{titlepage}


\bigskip

\bigskip 

\bigskip 

\bigskip 

\vspace{3\baselineskip}

\begin{center}
{\bf \LARGE 
Thermal dark matter abundance under non-standard macroscopic conditions in the early universe  }

\bigskip

\bigskip

{\bf Archil Kobakhidze$^{\rm a}$, Michael A. Schmidt$^{\rm b}$ and Matthew Talia$^{\rm c}$ \\ }

\smallskip

{ \small \it 
$^{\rm a}$School of Physics, The University of Sydney, NSW 2006, Australia \\
$^{\rm b}$School of Physics, The University of New South Wales, Sydney, NSW 2052, Australia \\
$^{\rm c}$Astrocent, Nicolaus Copernicus Astronomical Center Polish Academy of Sciences,
ul. Bartycka 18, 00-716 Warsaw, Poland}

\bigskip
 
\bigskip

\bigskip

{\large \bf Abstract}

\end{center}
\noindent 
The standard theoretical estimation of the thermal dark matter abundance may be
significantly altered if properties of dark matter particles in the early
universe and at the present cosmological epoch differ. This may happen if, e.g., a
cosmological phase existed in the early universe during which dark matter
particles were temporarily unstable and their abundance was reduced through
their decays. We argue that a large class of microscopic theories which are
rejected due to the dark matter overproduction, may actually be viable theories
if certain macroscopic conditions were satisfied in the early universe. We
explicitly demonstrate our mechanism within the minimal supersymmetric standard
model with the bino-like lightest supersymmetric particle being a
phenomenologically viable dark matter candidate under the condition that the
early universe carried a global R-charge which induced the instability phase. 

 \end{titlepage}

\section{Introduction}

The relation between the production of dark mater particles as thermal relics within the hot big bang cosmology and their interactions with the ordinary luminous matter is a theoretically appealing aspect of the thermal dark matter paradigm. The most prominent example of the thermal dark matter is the dark matter  comprising of stable neutral particles of mass $\sim \mathcal{ O}(100 \text{ GeV})$ interacting with the known standard model particles with the strength of weak interactions, the so-called WIMPs (weakly interacting massive particles). WIMPs, being once in the early universe in chemical equilibrium with the ordinary matter, would populate today’s universe in abundance that is comparable to the observed dark matter abundance. The ongoing direct dark matter detection experiments, however, have found no evidence for the WIMP dark matter so far and thus impose strong constraints on the WIMP interaction cross section \cite{Akerib:2016vxi,Cui:2017nnn, Aprile:2017iyp}.

Within the standard WIMP paradigm, a small WIMP cross section inevitably results in the earlier WIMP decoupling from the primordial thermal plasma which leads to its excessive abundance. In fact, many well-motivated theoretical models predict overabundant dark matter. For example, the bino-like dark matter within the minimal supersymmetric standard model (MSSM) is overabundant for most of the parameter area, except the case when the bino is nearly degenerate in mass with the sleptons~\cite{Ellis:1999mm,Gomez:1999dk} or light scalar top quarks~\cite{Boehm:1999bj}.
A small dark matter abundance can also be obtained in the A-pole region with $2m_\chi\simeq m_A$, 
where bino dark matter resonantly annihilates via a pseudoscalar $A$
\cite{Djouadi:2001yk}.

The undesired consequence of the early WIMP decoupling can be avoided if
properties of WIMP in the early universe differ from the properties at the
current cosmological epoch. In Refs. \cite{Baker:2016xzo} and
\cite{Kobakhidze:2017ini}, a new cosmological phase has been postulated, during
which WIMPs become temporarily unstable and its abundance is reduced through
its decays. The instability phase in the previous works was attributed to the
microscopic  physics with additional fields and interactions, which make the
original dark matter models somewhat contrived. In the current work, we would
like to propose an alternative scenario, which relies entirely on the
macroscopic conditions in the early universe and leaves microscopic physics
unaltered.  

The key idea behind our mechanism is the following. Imagine the early universe
carries a non-zero  global (approximately) conserved charge which is associated
with a symmetry that ensures dark matter stability in the microscopic theory.
For charge densities above some critical value, it becomes energetically more
favourable to dump the charge into the vacuum rather than keeping it in the
form of thermal excitations (particles). Such spontaneous rearrangement of the
vacuum induces processes with the global charge non-conservation, such as dark
matter decays. If these processes proceed in equilibrium they will tend to
wash-out the global charge and eventually end the instability phase, during
which the dark matter abundance has been reduced down to the acceptable level.
The instability phase then is followed by the standard cosmological evolution.

The mechanism outlined above can be applied to a wide class of thermal (and non-thermal) dark matter models.  In what follows we explicitly demonstrate this within the MSSM with a
bino-like dark matter, which is stable because it is the lightest R-parity odd particle.  In
particular, we will argue that if the early universe carried sufficiently large
global R-charge, a phase transition would occur and both R-parity and R-charge violating processes are
turned on in the new phase. During this phase bino dark matter abundance is
reduced through the bino decays, while R-charge gets also washed out by the
same equilibrium R-violating processes. Once the R-charge density becomes
smaller than some critical density, the R-parity is restored and the instability phase ends and the standard cosmological evolution with the reduced abundance of binos takes over.

Large charge asymmetries can be motivated by the solutions of topological defects problems~\cite{Bajc:1997ky,Bajc:1998rd}. In particular, Ref.~\cite{Bajc:1998rd} pointed out that R-charge may provide a simple way out of the monopole problem. Large charge asymmetries may also alter the nature of the phase transition~\cite{Schwarz:2009ii}.
 
The
paper is organised as follows. In the next section we introduce the model and
discuss the conditions for the existence of the instability phase. The
analytical estimation of dark matter abundance is given in section 3 and section 4 is reserved for conclusions.    

\section{Phase transitions in MSSM with the global R-charge}
We consider the sequestered scalar sector of MSSM which contains the lightest Higgs doublet field $\phi$ and a single generation of the left-handed slepton field  $\tilde{L}$. The tree-level potential at zero temperature reads: 
\begin{equation}
V_0 = -\frac{m^2_h}{2} |\phi|^2 + m^2_{\tilde{L}} |\tilde{L}|^2 + \frac{m^2_Z}{8v^2} \left( \frac{2m_h}{m_Z} |\phi|^2 + 2|\tilde{L}|^2 \right)^2\;.
\label{1}
\end{equation}
Here, $m_h\approx 125$ GeV is the Higgs boson mass, $m^2_{\tilde{L}}$ is the soft supersymmetry breaking sneutrino mass parameter and the final term comes from the  $D$-term contribution. All other additional particles of MSSM (except for bino dark matter) are assumed to be heavy and are irrelevant for our discussion. This low energy theory exhibits a global $U(1)$ symmetry under the transformations:  
\begin{align}
	\phi  &\rightarrow  {\rm e}^{i\alpha} \phi,&
	\tilde{L} &\rightarrow {\rm e}^{i\alpha} \tilde{L},&
	E_L^c  &\rightarrow e^{-i\alpha} E^c_L, \\
	Q &\rightarrow {\rm e}^{-i2\alpha /3}Q,&
	U^c_L &\rightarrow  {\rm e}^{-i\alpha /3} U^c_L ,&
	D^c_L &\rightarrow {\rm e}^{-i\alpha /3} D^c_L,
\label{2}
\end{align}
where $U^c_L, D^c_L$ and $E^c_L$ denote 3 generations of weak isospin-singlet up, down and charged lepton charge-conjugate left-chiral fields, respectively, while $Q=(u,d)_L$ denote 3 generations of weak isospin-doublet left-chiral quark fields.  In fact, this symmetry is a global R-symmetry of the total MSSM Lagrangian (except of the soft supersymmetry breaking trilinear scalar terms) with the R-charge assignment \cite{Dvali:1994qf} for the chiral superfields\footnote{Note that this global R-symmetry holds even when renormalizable R-parity violating terms are included.  The symmetry is explicitly broken by the soft supersymmetry breaking trilinear scalar terms. We assume that this breaking is small and cosmologically irrelevant.}:
\begin{equation}
	R_{\phi_i}=R_{L}=+1,
	~~~R_{Q}=1/3,
	~~~R_{u^c}=R_{d^c}=2/3,
	~~~R_{e^c}=0.
\label{3}
\end{equation}

In the cosmological context, our key assumption is that the primordial plasma at early times carried non-zero global R-charge. It is well-known \cite{Haber:1981ts, Benson:1991nj} that for some critical values of the non-zero global charge the symmetric vacuum state becomes unstable and is spontaneously rearranged into a new asymmetric vacuum state which carries a non-zero global charge. Utilising  the results of Ref. \cite{Bajc:1999he}, we calculate a correction to the scalar potential (\ref{1}) due to the non-zero chemical potential associated with the global R-charge and finite temperature corrections in the high temperature approximation: 
\begin{equation}
V=V_0+V_T+V_{n_R}~,
\label{4}
\end{equation}
where
\begin{eqnarray}
V_T&=&\alpha_hT^2 |\phi|^2 +\alpha_{\tilde L}T^2|\tilde L|^2, \\
\alpha_h &=&\frac{1}{8v^2}\left(4m_W^2+2m_Z^2+4m_t^2+m_h^2+\frac{2}{3}m_Zm_h\right)\approx 0.383~, \\
\alpha_{\tilde L} &=&\frac{1}{8v^2}\left(4m_W^2+4m_Z^2+\frac{1}{3}m_Zm_h\right)\approx 0.129~,
\label{5}
\end{eqnarray}
\begin{eqnarray}
V_{n_R}&=&\frac{1}{2} \frac{n^2_R}{\mathcal{M}} \nonumber \\
\mathcal{M}&=&\frac{T^2}{6}\left( \sum_{i={\rm fermions}} R^2_i + 2 \sum_{i={\rm bosons}} R^2_i\right) + 2  \left(|\phi|^2+|\tilde L|^2\right) \nonumber \\
&=&\frac{17T^2}{6} + 2  \left(|\phi|^2+|\tilde L|^2\right)
\label{6}
\end{eqnarray}
where $R_i$ are R-charges for the relativistic fermionic and bosonic states and $n_R$ is the R-charge density. Here we assume that all the supersymmetric particles are non-relativistic and their thermal corrections are neglected. For temperatures above the mass thresholds for supersymmetric particles, which we collectively denote $M_{SUSY}$, the temperature dependent terms become larger due to the positive contribution from additional relativistic species.

At this point, it is convenient to introduce the thermal masses,
\begin{eqnarray}
m^2_h(T)&=&-m^2_h+2\alpha_h T^2~, \\
m^2_{\tilde{L}}(T)&=&m^2_{\tilde{L}}+\alpha_{\tilde{L}} T^2~,
\end{eqnarray}
and rewrite the finite density and temperature effective potential as:
\begin{equation}
V=\frac{m_h^2(T)}{4}v_h^2+\frac{m^2_{\tilde{L}}(T)}{2}v_{\tilde{\nu}}^2+\frac{3n_R^2}{17T^2+6v_{\tilde{\nu}}^2+6v_h^2}~,
\label{eff_pot}
\end{equation}
where $v_{\tilde{\nu}}$ and $v_h$ expectation value for sneutrino and the Higgs, respectively at finite R-charge and temperature. We have omitted the terms with $v_{\tilde{\nu}}$ and $v_h$ in powers higher than 2, as we are in the regime where they are sub-dominant compared to the terms in (\ref{eff_pot}). The effect of the finite density of R-charge is that above some critical density $n_R^c$ it becomes more favourable to dump the excess of R-charge into the ground state through the sneutrino and/or Higgs condensates.  In particular, the minimisation of (\ref{eff_pot}) reveals that for the R-charge densities,
	\begin{equation}
		n_R\gtrsim n_R^c\simeq \frac{17}{6}T^2\, {\rm max}\left\lbrace \sqrt{m^2_{\tilde{L}}(T)}, \frac12\left(\sqrt{m^2_{\tilde{L}}(T)}+\sqrt{\frac{m_h^2(T)}{2}}\right)\right\rbrace~, 
\label{critical}
\end{equation}
the ground state configuration is the one with non-zero sneutrino expectation value, 
	  \begin{equation}
v^2_{\tilde{\nu}}\simeq \frac{n_R}{\sqrt{m^2_{\tilde{L}}(T)}}-\frac{17}{6}T^2\approx \frac{n_R}{m_{\tilde{L}}}-\frac{17}{6}T^2,  
\label{vev}
\end{equation}
and zero expectation value for the Higgs field, $v_h=0$. Here we assumed $m^2_{\tilde{L}}(T)> m^2_h(T)$, which is naturally satisfied, and that sneutrino is non-relativistic. 
 
 Hence, the following picture emerges. At very high temperatures $T> M_{SUSY}$ supersymmetric and standard model particles are in chemical and thermal equilibrium and carry collectively a net R-charge density. As temperature cools below $M_{SUSY}$ the sneutrino expectation value turns on and the instability phase starts, during which depletion of the bino dark matter occurs. Because $n_R$ is dominated by the relativistic species (right-handed charged leptons, quarks and the Higgs) it is diluted faster (as $\propto T^3$) due to the cosmological expansion than the density on the right hand side of the inequality (\ref{critical}), which decreases as $\propto m_{\tilde L}T^2$. Therefore, at temperature
 \begin{equation}
	 T_f\simeq \frac{255}{(2\pi)^2 g_{*}}\frac{m_{\tilde{L}}}{Y_R^0}~,
 \label{end}
\end{equation}   
the inequality (\ref{critical}) fails and sneutrino expectation value becomes zero, ending the instability phase. In (\ref{end}), we introduced the R-charge yield, $Y_{R}=n_R/s$, where $s$ is the entropy density, together with its initial value $Y_R^0$. Since no substantial production of entropy is assumed and the depletion of the R-charge during the instability phase is inefficient (see the discussion below),  $Y_{R}$ is essentially constant and equal to the initial yield $Y_R^0$, which is one of the key input parameters in our model. With this ingredient we can now proceed to estimate the bino dark matter relic density.


\section{Depopulation of the bino dark matter}
\label{sec:depopulation}

\textbf{Standard case:}
In the absence of coannihilation processes, the dominant bino dark matter annihilation channels are via t-channel slepton exchange. In this context, this is mainly mediated by the light left-handed slepton doublet $\tilde L$ due to the assumed mass difference with the right-handed slepton. In any case the bino annihilation cross section is too small to keep bino dark matter sufficiently long in equilibrium and thus the naive prediction for the bino dark matter abundance is much larger compared to the measured value, $\Omega_{\text{Planck}}h^2 = 0.12 \pm 0.001$~\cite{Planck2018}.



\noindent\textbf{With decays:}
During the instability phase $x \in [m_{\chi}/M_{SUSY},x_f=m_{\chi}/T_f]$ the neutralino ceases to be a stable particle. More specially, condensation of the sneutrino field leads to a spontaneous breaking
of R-parity and to a mixing of neutralinos and neutrinos. Through this mixing the neutralino LSP decays into standard model particles, the dominant channels being 2-body decays $\chi \rightarrow Z\nu', W^{\pm}\ell^{\mp}$, where $m_{\chi} > m_Z(m_W) + m_{\nu}(m_{\ell})$. The mass matrix for bino-neutrino mixing is given by:
\begin{equation}
	\hat{M}_{\chi \nu}=\left(\begin{array}{ccc}
			0 &  g^{\prime} \frac{v_{\tilde \nu}(T)}{\sqrt{2}} \\ 
		g^{\prime} \frac{v_{\tilde \nu}(T)}{\sqrt{2}}  & {M_{1}}\end{array}\right)
\end{equation}
and thus $\sin\beta=\frac{g^\prime v_{\tilde\nu(T)}}{\sqrt{2} m_\chi}$
where we neglected the neutrino mass and the sneutrino thermal VEV is defined in Eq.~\eqref{vev}, which we further approximate for $n_R\gg m_{\tilde L}T^2$. The bino mass eigenstate then is,
\begin{eqnarray} \chi & \simeq & \tilde{B} \cos \beta -\nu \sin \beta  \nonumber \\ \sin \beta & \simeq & \frac{ g^{\prime} n_{R}^{1 / 2}}{m_{\chi}\left|m_{\tilde{L}}\right|^{1 / 2}}
\end{eqnarray}
with the mass $m_\chi \approx M_1$ (bino-like neutralino). With these ingredients we compute the decay width as~\cite{Kobakhidze:2017ini}: 
	\begin{equation}
		\Gamma_\chi = \frac{3g^{\prime2} m_\chi}{32\pi}~.
	\end{equation}
	In the given approximation, the sneutrino VEV cancels in the expression of $\Gamma_\chi$ (see App.~A in Ref.~\cite{Kobakhidze:2017ini} for details.), and thus the decay width is independent of temperature and as well does not explicitly depends on the R-charge density $n_R$. One must keep in mind, however, that as $n_R\to 0$, the sneutrino VEV tends to zero as well and the instability phase cease to exist.  

To calculate the dark matter yield we follow the discussion in Ref.~\cite{Kobakhidze:2017ini}. We assume that during the instability phase up until $x=x_f$ ($\Gamma_\chi(x_f) x_f^2 > H_\chi$) the neutralino decay rate dominates over the expansion rate and thus neutralino yield is close to its equilibrium abundance. This is satisfied for typical values and does not pose any constraint for TeV-scale neutralino dark matter. The dark matter yield today then is computed as:   
\begin{equation}
	Y_{\chi,0} \approx \frac{Y_\chi^{(eq)}(x_f) x_f}{Y_\chi^{(eq)}(x_f) s_\chi \langle\sigma v\rangle/H_\chi + x_f} \approx Y_\chi^{(eq)}(x_f)~, 
	\label{yield}
\end{equation}
where  $Y^{(eq)}_{\chi}(x) = \frac{45}{2 \pi^{4}}\left(\frac{\pi}{8}\right)^{1 / 2} \frac{g_{\chi}}{g_{*}} x^{3 / 2} \mathrm{e}^{-x}$ is the equilibrium number density of the non-relativistic bino dark matter, and $s_{\chi} = \left(2 \pi^{2} / 45\right) g_{*} m_{\chi}^{3}$ and $H_{\chi}=\left(\pi^{2} g_{*} / 90\right) \frac{m_{\chi}^{2}}{M_{P}}$ denote the entropy density and Hubble rate at $x=1$.	The thermally-averaged annihilation cross section $\langle\sigma v\rangle$ can be approximated by~\cite{ArkaniHamed:2006mb}
\begin{equation}
\left\langle\sigma v\right\rangle \approx\frac{g^{4} \tan ^{4} \theta_{W}\, r\left(1+r^{2}\right)}{16\pi m_{\tilde{L}}^{2} \,x\,(1+r)^{4}},  \quad r \equiv \frac{m_\chi^{2}}{m_{\tilde{L}}^{2}}\;. 
\end{equation}
The second approximate equality in Eq.~\eqref{yield} follows, because the first term in the denominator can generally be neglected for bino dark matter in the absence of coannihilation due to the Boltzmann suppression of $\chi$. 
Hence the dark matter yield today is essentially given by the yield at the end of the instability phase, $x=x_f$.   

Finally, we can recast (\ref{yield}) into the present-day abundance of neutralinos as:
\begin{equation}
\Omega_{\chi} = \frac{m_{\chi} Y_{\chi,0} s_{0}}{3H^2_{0}M^2_P}\, ,
\end{equation}
where $s_0 = 2891.2\,\text{cm}^{-3},$ is today's entropy density and $H_0 = 100 h\,\text{km s}^{-1} \text{Mpc}^{-1}$ 
$(h = 0.673)$ today's Hubble rate. This leads to
	\begin{equation}
		\Omega_{\chi}h^2 = 0.12 \left(\frac{m_\chi}{3.0\mathrm{TeV}}\right) \left(\frac{Y_R^0 m_{\chi}}{1.8m_{\tilde{L}}}\right)^{3/2} \exp\left(-\frac{424 \pi^2}{255}\frac{ Y_R^0m_{\chi}}{1.8m_{\tilde{L}}}\right)\,.
\label{final}
\end{equation}
Note, the final dark matter abundance in (\ref{final}) is essentially defined by the initial R-charge density, $Y^0_R$. It also does not depend explicitly on the decay rate $\Gamma_{\chi}$. 

\begin{figure}\centering
	\includegraphics[width=0.7\linewidth]{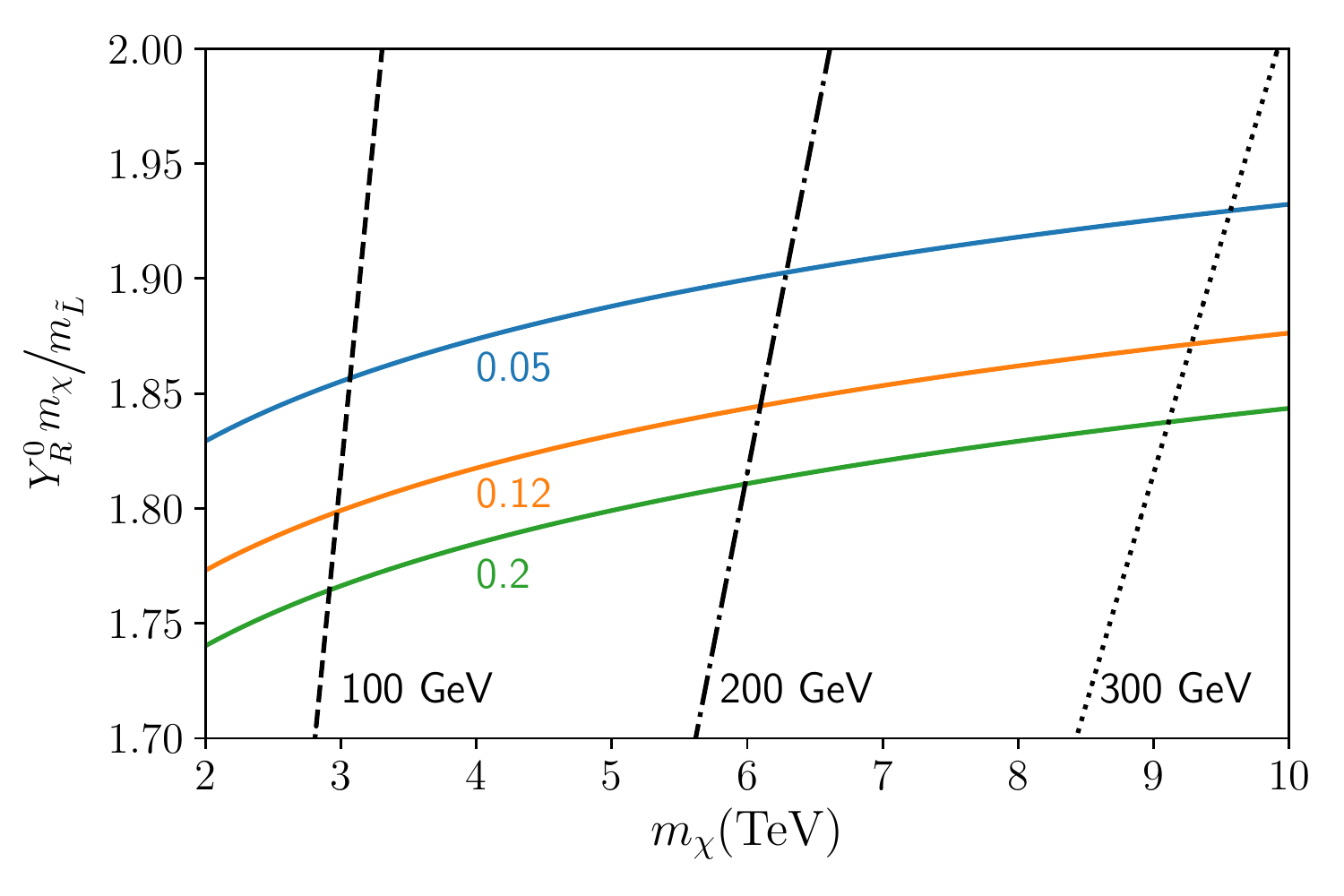}
	\caption{Contour plot of the DM abundance
	$\Omega h^2$ as a function of the DM mass
$m_{\chi}$ in TeV and the initial $R$-charge
normalized by the mass ratio of the DM mass to the
sneutrino mass, $Y_R^0 m_\chi/m_{\tilde L}$.  
The label on each coloured solid curve denotes the value of $\Omega h^2$. The dashed, dot-dashed, and dotted lines indicate the temperature when the instability phase ends $T_f$.}
	\label{fig}
\end{figure}

The neutralino depopulation through its decays may
explain the observed dark matter density as
demonstrated in Fig.~\ref{fig}. The final abundance
is essentially defined by a few parameters. One is
a macroscopic parameter $Y_R^{0}$, the initial
value of R-charge yield. Two others are microscopic
parameters of the theory: the dark matter mass
$m_{\chi}$ and dark matter mass-to-sneutrino mass
ratio, $m_{\chi}/m_{\tilde L}$. The later mass
parameter, $m_{\tilde L}$, defines the scale of the
spontaneous R-charge breaking, which triggers the
instability phase. One can see from Fig.~\ref{fig}
that a right amount of the initial R-charge
asymmetry, $Y_R^{0}$, predicts correct abundance of
dark matter (middle curve) for a large range of
masses, while smaller (large) values of $Y_R^{0}$
result in over(under)-abundance of dark matter.
Hence, the dark matter abundance in our scenario
crucially depends on the macroscopic parameter of
R-charge yield. 

In this scenario, the left-handed
	sneutrinos and sleptons are generally
	light. For the correct dark matter abundance their mass is directly related to
	the dark matter mass $m_\chi$ and the initial
	R-charge density $Y_R^0$, $m_{\tilde L} \simeq 0.56
	Y_R^0 m_\chi$, and thus naturally in the
	multi-TeV range. This opens up the
	possibility to search for sleptons, which
	promptly decay to a charged lepton and the
	lightest neutralino. The standard LHC
	searches apply, which currently are not
	sensitive to the multi-TeV region. The most
	recent analysis of slepton pair
	production~\cite{ATLAS:2019cfv} has been
	performed by the ATLAS experiment. It
	probes slepton masses up to 700 GeV, but is
	not sensitive to the multi-TeV range.

	Sleptons contribute to dark matter nucleus scattering at loop-level via the anapole operator $\mathcal{L}=c_A\bar\chi\gamma^\mu\gamma^5\chi J_\mu^{em}$. 
See Refs.~\cite{Berlin:2015njh,Herrero-Garcia:2018koq} for detailed discussions of dark matter direct detection at loop level.
For light sleptons this contribution dominates over the tree-level Higgs exchange and results in the bino-proton scattering cross section~\cite{Berlin:2015njh} $\sigma=2\alpha m_p E_R c_A^2$ in terms of the proton mass $m_p$, the fine structure constant $\alpha$, the recoil energy $E_R$, and the Wilson coefficient of the anapole operator squared, $c_A^2\sim \alpha^3/m_{\tilde L}^4$. It is suppressed due to its dependence on the recoil energy $E_R$ 
and thus current direct detection experiments do not pose any constraint, as it is illustrated in Fig.~13 of Ref.~\cite{Berlin:2015njh}. 

This mechanism can be applied to gravitino dark matter. As it has been shown the finite R-charge induces temporally R-parity violation. This allows the gravitino to decay to SM particles, e.g.~decaying to a photon and a neutrino, see e.g.~Ref.~\cite{Dudas:2018npp}. More generally the cosmological gravitino problem~\cite{Moroi:1993mb} can be solved via this mechanism. It may be interesting to study this possibility in more detail.

\section{Conclusions}
\label{sec:conclusion}
The thermal dark matter abundance besides its
microphysical properties, critically depends on the
details of cosmological evolution. The latter can
be altered through macroscopic conditions in the
early universe such that dark matter properties in
the early universe and at present may substantially
differ. In this paper, we have explored this
scenario to argue that a  large class of models
which predict incorrect dark matter may actually be
phenomenologically viable under non-standard
macroscopic conditions in the early universe. 

The particular example illustrating the above point
is the bino-like dark matter within the
supersymmetric extension of the Standard Model,
which is known to be overabundant for the most of
the range of parameters. We have demonstrated that
in the presence of large enough global R-charge
asymmetry in the early universe, the phase
transition occurs where both R-charge and R-parity
are spontaneously broken followed by restoration of
those symmetries at lower temperatures. This
defines a phase in the cosmological evolution of
bino dark matter during which it is unstable.  The
abundance of bino dark matter is then reduced due
to its decays during the instability phase. As a
result, the presently observed dark abundance can
be attained in the model which would be considered
phenomenologically unacceptable within the standard
cosmological framework. The resulting dark matter
abundance in our scenario  crucially depends on the
macroscopic R-charge density in the early universe. 

As a final comment, we note that relevance of our
mechanism for dark matter depopulation goes beyond
the particular supersymmetric dark matter model
discussed,  and can be applied to a wide range of
dark matter models including those with non-thermal
dark matter production. The mechanism is able to reduce a preexisting dark matter abundance, irrespective of the production mechanism, as long as there is an instability phase.
 
\paragraph{Acknowledgements} 
The work of AK was partially supported by Shota Rustaveli National Science Foundation of Georgia (SRNSFG) [DI-18-335/New Theoretical Models for Dark Matter Exploration]. MT is supported by the grant "AstroCeNT: Particle Astrophysics Science and Technology Centre" carried out within the International Research Agendas programme of the Foundation for Polish Science financed by the European Union under the European Regional Development Fund.


\end{document}